\documentstyle[12pt]{article}

\newtheorem{algorithm}{PDP Algorithm}
\newtheorem{ralgorithm}{Relativistic PDP Algorithm}
\def\lqq{\lq\lq}
\def\rqq{\rq\rq}
\newcommand{\ha}{H_\alpha}
\newcommand{\gab}{g_{\alpha\beta}}
\newcommand{\gba}{g_{\beta\alpha}}
\newcommand{\la}{\Lambda_\alpha}

\newcommand{\be}{\begin{equation}}
\newcommand{\ee}{\end{equation}}

\newcommand{\ra}{\rho_\alpha}

\newcommand{\aaa}{A_\alpha}
\begin{document}

\begin{titlepage}
\today          \hfill 
\begin{center}
%\hfill    Preptint No \\
\vskip .2in
{\large \bf EEQT \\ A Way Out of the Quantum Trap}
%\footnote{Thanks text}
\vskip .50in
\vskip .2in

Ph.~Blanchard${}^\flat$ \ and\ A.~Jadczyk${}^\sharp$\footnote{
e-mail: ajad@physik.uni-bielefeld.de}

{\em ${}^\flat$ Faculty of Physics and BiBoS,
University of Bielefeld\\
Universit\"atstr. 25,
D-33615 Bielefeld}
\end{center}

\vskip .2in

\begin{abstract} 
We review Event Enhanced Quantum Theory (EEQT). In Section 1 we address the
question "Is Quantum Theory the Last Word". In particular we respond to
some of challenging staments of H.P. Stapp. We also discuss a possible
future of the quantum paradigm - see also Section 5. 
In Section 2 we give a short sketch
of EEQT. Examples are given in Section 3. Section 3.3 discusses
a completely new phenomenon - chaos and fractal-like phenomena
caused by a simultaneous "measurement" of several non-commuting
observables. In Section 4 we answer "Frequently Asked Questions"
concerning EEQT - mostly coming from referees of our publications.
Summary and conclussions are in Section 6.

\end{abstract}

\end{titlepage}
\newpage
\section{Introduction:\\ Is Quantum Theory the Last Word?}
Sixty years after the famous debate on the nature of Reality between Niels Bohr and
Albert Einstein, questions central to their debate are the subject of
fascinating experiments.
Is Quantum Theory the last word?  Have we plumbed the depth and spanned
the breadth of scientific inquiry and found that there just simply
is no more? Have we come to the "end of the line?" In ancient times, the
number of things that could be known was limited and it WAS possible for a
single person to know them all.  Even as recently as 200 years ago, our range
of knowledge was severely circumscribed by our assumptions about the world around us.  In the previous century, daring thinkers and observers expanded our understanding of the world in which we live to such a fantastic degree that man's technological progress in the past 100 years has surpassed the previous 2,000 years added all together.  Does this acceleration indicate that the end is near?  Or, shall we compare such claims to the story about the examiner in the U.S. Patent Office who, at the beginning of this century, suggested that the Patent Office be closed since "everything has already been invented."  Well, it clearly wasn't the end in 1901, but perhaps we are approaching it now in  the year 2001? Perhaps science, as we know it, has become obsolete because it has explained everything... there is no more "to invent."  Does this sound fantastic?  Well, one certainly gets this impression when reading some of the recent papers on the subject - a remarkable example being H. P. Stapp's contribution to the X-th Max Born "Quantum Future" Symposium \cite{born}, entitled "Quantum Ontology and Mind-Matter Synthesis."
\cite{stapp}\\

We wish to address Stapp's main theses later on in this section, but before
we do, we would like to address the question that certainly deserves an answer if we want to be honest with our audience, to wit: IF quantum theory is NOT The Last Word, if there IS a future, what can it be? It is clear that our answer to this question must be, at present, based on speculation. But, as part of the tradition, once in a while, scientists do speculate! \cite{good} \\
Our answer, given speculatively remember, is that we believe that quantum theory is an effective and powerful theory of measurements. But, we also believe that it deals only with a particular aspect of reality and that other approaches
can, and shall, and must, give us a deeper insight into the complex weavings
of Nature.

When we speak about Quantum Theory, we mean its standard and orthodox version which is clearly a linear theory. It is very good for making predictions based on incomplete information. But, it is not suitable for explaining what really happens. It explains the objects and phenomena of our experience, but does NOT explain the underlying reality which we do not experience directly. As Alan Turing succintly stated: "prediction is linear, description is non-linear".  There will be more on this subject in Sect. 6, where we will describe in more details a possible scenario for future developments in quantum theory - a "Quantum Future" that we believe is valid and deserving of serious work,  even if only to see if Stapp's dark vision of what lies ahead in science is what we must prepare ourselves to face.

Stapp states: "...I propose to break away from the cautious stance of the  founders of quantum theory, and build  a  theory of reality by taking seriously what the incredible accuracy of the predictions of the formalism seems to proclaim, namely that nature is best understood as  being built around knowings that enjoy the mathematical properties ascribed to them by  quantum theory."  According to Stapp, reality should be "recognized to  be  knowledge, rather  than substantive matter... "  

Stapp is not the first one to propagate the view that nothing
"out there" really exists. Bishop Berkeley, as we know, proclaimed the
very same idea long ago when he said: "I have no reason for believing the existence of matter." But Stapp is supporting this view by the
practical success of quantum mechanics and failure of "all of our efforts to rid
physics of this vile contamination by mind, which quantum theory
presses upon us."

Well, let us just point out that the success of quantum theory is not that overhelming!

Meaningless infinities of relativistic quantum field
theory tell us that something is seriously wrong with our theoretical assumptions.  In our opinion, the value of a theory consists not in that it can explain the technique by which the fabric is woven on the loom of Nature, but that it can explain the patterns of the weaving, the Weaver and perhaps the motivations behind the weaving.  

Facts cannot be understood by being crafted into a summary or a formula - they can only be understood by being explained.  And, understanding is not the same as "knowing." Quantum Theory, as any other theory, has a finite region of validity - when
attempts are made to apply it beyond these limits - we get either nonsense
or no answer at all. Quantum theory, in its orthodox version, cannot even
be applied to an individual system - like the Universe we live in and experience.   We want to discover "why" in addition to "what" regarding the order of the universe in which we find ourselves.  We wish to discover why "this" MUST be so, rather than "that;" why Nature does what she does and how.  We want to uncover and understand the Laws of Nature, not just the "rules of thumb."

Stapp knows this all too well, but apparently he has lost faith that a better
theory can be forthcoming, even if only one small step at a time. Einstein failed, Bohm failed (because his "model has  not  been   consistently  extended  to the relativistic case of quantum  electrodynamics, or to quantum chromodynamics, which are our premiere quantum theories."), Stapp himself, many years ago,  tried to advance his own "theory of events" - unsuccessfully it seems.  And  now, apparently, he is convinced that there is no way out. If he could not do it, it cannot be done.  Must it be so?  Can it not be that a better explanation is the one that leads to improvements in techniques and concepts and structure?\\
In his paper \lqq The Philosophy of Experiment\rqq\,  E. 
Schr\"odinger \cite{schrod55} 
wrote:
\begin{quotation}
\lqq  The new science (q.m.) arrogates the right to bully our whole
philosophical outlook. It is pretended that refined measurements which
lend themselves to easy discussions by the quantum mechanical formalism
could actually be made. (...) Actual measurements on single individual
systems are never discussed in this fundamental way, because the theory
is not fit for it.(...) We are also supposed to admit that the extent of
what is, or might be, observed coincides exactly with what the quantum
mechanics is pleased to call observable.\rqq
\end{quotation}

As we have stressed elsewhere \cite{ja94c}
\begin{quote}
J.S. Bell \cite{bell90,bell89} deplored the misleading use of the term
\lqq measurement\rqq\,  in quantum theory. He opted for banning this word 
from our quantum vocabulary, together with other vague terms such as
\lqq macroscopic\rqq, \lqq microscopic\rqq, \lqq observable\rqq\,  and 
several others. He suggested that we ought to replace the term \lqq measurement\rqq\,  with that of \lqq experiment\rqq, and
also not to even speak of \lqq observables\rqq\,  (the things that seem to call 
for an
\lqq observer\rqq) but to introduce instead the concept of \lqq 
beables\rqq  - the
things that objectively \lqq happen--to--be (or
not--to--be)\rqq.\footnote{Calling observables \lqq observables\rqq\,  can 
be, however, justified in the event-enhanced formalism that we are outlining
here.} 

But there is no place for \lqq events\rqq\,  or for \lqq 
beables\rqq\, in ordinary
quantum theory. That is because each \lqq event\rqq\,  must have three
characteristic features:
\begin{itemize}
\item it is classical,
\item it is discrete,
\item it is irreversible.
\end{itemize}
If just one of these three features is relaxed, then what we have is not
an \lqq event\rqq.\\
It must be {\sl classical}, because it must obey the classical
\lqq yes-no\rqq\,  logic; it must never be in a \lqq superposition\rqq\, of 
it being "happened and/or unhappened." Otherwise it would not be an event.\\
It must be {\sl discrete}. It must happen wholly. An event that \lqq 
approximately\rqq\, happened is not an event at all.\\
It must be {\sl irreversible}, because it can not be made to be \lqq undone\rqq. 
This feature distinguishes real events from the \lqq virtual\rqq\,  ones. Once
something has happened -- it has happened at a certain instant of time. It must
have left a trace. Even if this trace can be erased, the very act of erasing will change the future -- not the past. Something else may
happen later, but it will be already a different event. We believe
that {\sl events, and nothing but events, are pushing forward the 
arrow of time}.
\end{quote}

German philosopher E. Bloch expressed the very same idea succintly: {\em Zeit ist nur dadurch, da\ss{} etwas geschieht und nur dort wo
etwas geschiecht.}
But Stapp will say: "where are these events if not in our minds alone?"
Our answer is: every energy transfer from one place to another is an
event. Do such energy transfers happen?  We speculate that they do. Our engineering and technology stands as proof. But then, one could inquire, where
precisely do we think these events happen?  We answer: they are localized neither in space nor in time. But, in our simplified mathematical models of reality we associate events with  particular pieces of our experimental setup, mainly with perceivable or recordable changes of macroscopic bodies.\footnote{A similar evolutionary view of Nature is pursued in
a recent series of papers by R. Haag \cite{haag90a,haag90b,haag96a,haag96b,haag98}. According to this view the Future does not
yet exist and is being continuously created, this creation being marked by events. That view
parallels ours, but we are more open towards a "dynamical many worlds algorithm" - cf. \cite{deutsch,wolf} and references therein} 

This is not to deny the existence or importance of "mental events" or "knowing," or the part that "mind" plays in the observing/measuring process - but these are only part of the answer.  We agree with Stapp in one (but only one) point: the orthodox quantum theory is about measurements rather than about the real world that is being measured. The orthodox quantum theory is about predictions based on "knowledge of the observer." But orthodox quantum theory is not the only theory in existence, and it grasps only a piece of what can be grasped.  It explains only part of the problem of how and why Nature weaves as she does.  It comes nowhere close, in our opinion, to explaining everything that can be understood.

If knowledge, this thing that Stapp considers so highly, is to continue to grow, then the depth and breadth of the theory must expand as well.  Stapp says: "This structure evolves the knowledge created by earlier knowings into the makings of later knowings" and "It is rather the knowings that are the basic irreducible units: they enter as entire units into a dynamic structure that carries forward the facts fixed by past knowings to produce the possibilities for future knowing"  which actually amounts to the same "clockwork theory" of classical physics only at a greatly reduced scale and with the locus of manifestation reversed!  One thing he says with which we agree up to a point: "Orthodox quantum theory is pragmatic; it is a practical tool based on human knowing."

But we seek to bridge the gap between "knowing" and understanding.  Event Enhanced Quantum Theory, or EEQT as we denote it -- cf. \cite{blaja93a,blaja95a,blaja95e,blaja95f} -- is a minimal
extension of quantum theory that accounts for events. It is a minimal
extension of quantum theory that unifies continuous evolution of
"wave function" with quantum jumps that accompany real world events.
We do not pretend that EEQT is a fundamental theory. It is semi-phenomenological
in its nature. But it shows that one can go beyond linear quantum theory,
that one can predict more than the standard formalism would allow,
that new questions can be asked, new horizons opened against the gloomy
fog  of the "nothing but knowings" landscape of future physics.

 In the eighties
fundamental concepts of quantum theory:
\begin{itemize}
\item[$\bullet$] wave particle duality
\item[$\bullet$] quantum state vectors
\item[$\bullet$] back--action in quantum measurement
\item[$\bullet$] uncertainty limits
\item[$\bullet$] Schr\"odinger cats (S. Haroche \cite{haroche})
\end{itemize}
gradually became accessible to experimentalists. The practical questions
of controlling fundamental quantum phenomena have surfaced in the domains of
quantum optics and applied physics. Quantum optics, in particular, has
a special fascinating flavor as it deals with
\begin{enumerate}
\item Squeezing
\item Quantum Non-Demolition
\item Quantum State Reconstruction
\item Cavity Quantum Electrodynamics; Quantum Optics of Single Atoms
 (S. Haroche\cite{haroche}, H. Walther\cite{walther})
\item Quantum Information:
\begin{itemize}
\item Quantum Cryptography (A. Ekert \cite{qubit0}, N. Gisin,...)
\item Quantum Computers (D. Deutsch, P.W. Shor, P. Zoller,...\cite{qubit})
\item Quantum Teleportation (F. De Martin, A. Zeilinger ) 
\end{itemize}
\item Spatial Quantum Structures
\end{enumerate}

New technologies and new experiments need a new theory that will allow for simulation of real--time behaviour of individual quantum systems communicating with external
control devices. EEQT is such a theory - it was created just for
this purpose.\\  
%%%%%%%%%%%%%
In Section 2 we will describe the mathematical formalism of EEQT and in
Section 3 we will list its main results, in particular
application of EEQT to the problem of relativistic quantum measurements.
The main point made there is: the decision mechanism for events in EEQT
is non-local in space. In a relativistic theory it must also be non-local in time. 
This implies that, once in a while,
the effect will precede its cause. We expect to see events that can
be interpreted in terms of superluminal propagation. The probabilistic
character of such a propagation, as described in our model, prevents
anti-telephone paradoxes from taking place.\\ It is to be noted that
superluminality is a question that continues to fascinate physicists
and laymen alike. Recently, with a new generation of tunneling time
experiments, it has become a laboratory experimental question. The
concept of tunneling time is well posed in EEQT - cf. \cite{muga97,
rusch}

It is also to be noted that our relativistic model lives in a five-dimensional space-time, with Schwinger-Fock "proper time" as the fifth coordinate, and also
that we are using indefinite-metric Hilbert space. This last property
does not contradict positive definiteness of probabilities in our model.

It should be stressed that in EEQT all the probablistic interpretations of
quantum theory are derived from the dynamics! In particular, it makes no
sense to ask the question "what would be a distribution of observed
values of an observable" without adding the appropriate
terms to the evolution equation. In this respect EEQT embodies in its
dynamics much more of the spoken philosophical language of Bohr and
Heisenberg, quoted so freely by Stapp, than Standard Quantum Theory. There is a price that we must pay for this: the dynamical equations of EEQT are harder to solve. But, on the other hand, EEQT makes it possible to analyze experimental situations that the
Standard Quantum Theory seems to exclude from its consideration -
like simultanous measurement of several noncommuting observables.
In this case, as explained in more detail in Section 3, measurement
results exhibit chaotic and fractal behaviour.

In Section 5 we will attempt to answer frequently asked questions and objections against EEQT.  In Section 6, we will sketch possible future developments of EEQT, while Section 7 will summarize our paper.

\section{EEQT -- mathematical formalism}

There are two levels of EEQT - the ensemble level and the individual level.
Let us consider first the ensemble level.\\
First of all, in EEQT, at that level, we use all the standard mathematical formalism of quantum
theory, but we extend it adding an extra parameter $\alpha .$ Thus
all quantum operators $A$ get an extra index $A_\alpha$, quantum
Hilbert space ${\cal H}$ is replaced by a family ${\cal H}_\alpha ,$
quantum state vectors $\psi$ are replaced by families $\psi_\alpha ,$
quantum Hamiltonian $H$ is replaced by a family $H_\alpha$ etc.\\ 
The parameter $\alpha$ is used to distinguish between macroscopically
different and non-superposable states of the universe. In the simplest
possible model we are interested only in describing a "yes-no" experiment
and we disregard any other parameter - in such a case $\alpha$ will have
only two values $0$ and $1$. Thus, in this case, we will need two Hilbert
spaces. This will be the case when we will deal with particle detectors. 
In a more realistic situation $\alpha$ will take values in a
multi-dimensional, perhaps even infinite-dimensional manifold - like, 
for instance, in a phase space of a tensor field. But even that may 
prove to be insufficient. When, for instance, EEQT is used as an engine
powering Everett-Wheeler many-world branching tree, in that case
alpha will also have to have the corresponding dynamical branching tree structure, where the space in which the parameter $\alpha$ takes 
values, grows and becomes more and more complex together with the growing complexity of the branching structure.\\ 
An event is, in our mathematical model, represented by a change of
$\alpha .$ This change is discontinuous, is a branching. Depending on
the situation this branching is accompanied by a more or less radical
change of physical parameters. Sometimes, like in the case of a phase 
transition in Bose-Einstein condensate, we will need to change the 
nature of the underlying Hilbert space representation. In other cases,
like in the case of a particle detector, the Hilbert spaces ${\cal H_0}$
and ${\cal H_1}$ will be indistinguishable copies of one standard
quantum Hilbert space ${\cal H}.$ \\
A second important point is this: time evolution of an individual quantum system is described by piecewise continuous function $t\mapsto\alpha(t)$,
$\psi(t)\in{\cal H}_\alpha(t)$, a trajectory of a piecewise deterministic
Markov process. The very concept and the theory of piecewise deterministic processes (in short: PDP) is not a part of the standard mathematical education, even for professional probabilists. But the point is that it is impossible to
unerstand the essence of EEQT without having even a rough idea about PDPs.\\
Originally EEQT was described in terms of a master equation for a coupled,
quantum+classical, system. Thus it was only applicable to ensambles - the question 
of how to describe individual systems was open. Then, after searching through
the mathematical literature, we found that in his 
monographs \cite{davmha1,davmha2}
dealing with stochastic control and optimization
M. H. A. Davis, having in mind mainly queuing and insurance models,
described a special class of piecewise deterministic
processes that fitted perfectly the needs of quantum 
measurement theory, and that reproduced the master equation postulated
originally by us in \cite{blaja93a}. \\
It took us another couple of years to show \cite{jakol95} that the special 
class of couplings between a
classical and quantum system leads to a unique piecewise deterministic
process with values on $E$-the pure state space of the total system.
That process consists of random jumps, accompanied by changes of a
classical state, interspersed by random periods of Schr\"odinger-type
deterministic evolution. The process, although mildly nonlinear in quantum 
wave function $\psi$, after averaging, 
recovers the original linear master equation for statistical states.\\

It should be stressed that in EEQT
the dynamics of the coupled total system which is being modelled
is described not only by a Hamiltonian ${\cal H}$, or better: not
only by an $\alpha$-- parametrized family of Hamiltonians $H_\alpha$,
but also by a doubly parametrized family of operators $\{g_{\beta\alpha}\}$,
where $g_{\beta\alpha}$ is a linear operator from ${\cal H}_\beta$
to ${\cal H}_\alpha$. While Hamiltonians must be essentially self-adjoint,
$g_{\beta\alpha}$ need not be such -- although in many cases, when information transfer and control is our concern (as in quantum computers), one wants them to be even positive operators (otherwise unnecessary entropy is created). This aspect of EEQT is rather difficult to accept 
for a newcomer, as the first question he will ask is "where do we take
these operators from?" Our answer, elaborated in more details in FAQ-s
- see Section 5 -- amounts to this: we find the correct operators
$g_{\alpha\beta}$ the same way we find the correct Hamiltonians:
by trial and error! Each new solved model is a lesson and, little
by little, we learn more and more, and we aspire for more. As already said,
more on this subject in our FAQs section.\\
It is to be noted that the time
evolution of statistical ensembles is due to the presence of $\{g_{\beta\alpha}\}$'s,
non-automorphic. The system, as a whole, is open. This is necessary, as we like
to emphasize: information (in this case: information gained by the classical
part) must be paid for with dissipation! The appropriate mathematical formalism
for discussing the ensemble level is that of completely positive semigroups, 
as discussed by Kossakowski et al. \cite{koss}, Lindblad \cite{lin}
and generalized so as to fit our purpose by Arveson \cite{arv} and 
Christensen \cite{chr}.\\
A general form of the linear master equation describing statistical evolution
of the coupled system is given by
\be
{\dot A}_\alpha=i[\ha,\aaa]+\sum_\beta \gba^\star
A_\beta \gba - {1\over2}\{\la,\aaa\},\label{eq:lioua}
\ee
\be
{\dot \rho}_\alpha=-i[\ha,\ra]+\sum_\beta \gab
\rho_\beta \gab^\star - {1\over2}\{\la,\ra\},\label{eq:liour}
\ee
where
\be
\la=\sum_\beta \gba^\star \gba.
\ee
The operators $\gab$ can be allowed to depend explicitly
on time. While the term with the Hamiltonian describes "dyna-mics",
that is exchange of forces, 
of the system, the term with $\gab$ describes its "bina-mics" -
that is exchange of "bits of information" between the quantum and the classical
subsytem.\\

As has been proven in \cite{jakol95} the above Liouville equation,
provided the diagonal terms $g_{\alpha\alpha}$ vanish, can be
considered as an average of a unique Markov process governing the
behavior of an individual system.
The real--time behavior of such an individual system is given by a
PDP process realize by the following 
non--linear and non--local, EEQT algorithm:
\begin{algorithm}
Suppose that at time $t_0$ the 
system is described by a quantum state vector $\psi_0$ and a classical 
state $\alpha$. 
Then choose a uniform random number $p\in [0,1]$, and proceed with
the continuous time evolution by solving the modified Schr\"odinger 
equation
$$
{\dot \psi_t}=(-i\ha dt-{1\over2}\la )\psi_t
$$
with the initial wave function $\psi_0$ until $t=t_1$, where $t_1$
is determined by
$$\int_{t_0}^{t_1} (\psi_t,\la \psi_t ) dt = p.$$ Then jump.
When jumping, change $\alpha\rightarrow 
\beta$ with probability 
$$p_{\alpha\rightarrow\beta}=
\Vert\gba\psi_{t_1}\Vert^2/(\psi_{t_1}\la,\psi_{t_1}),$$ and change 
$$\psi_{t_1}\rightarrow\psi_1=\gba\psi_{t_1}/
\Vert\gba\psi_{t_1}\Vert.$$ 
Repeat the steps replacing 
$t_0,\psi_0,\alpha$ with $t_1,\psi_1,beta$.
\end{algorithm}
The algorithm is non--linear, because it involves repeated normalizations.
It is non--local because it needs repeated computing of the norms - they
involve space--integrations. It is to be noted that PDP processes are more general
than the popular diffusion processes. In fact, every diffusion process can be
obtained as a limit of a family of PDP processes.

\section{EEQT -- a few examples}
\subsection{}
Let us consider the simplest case: that of a two--state classical system. We
call its two states "off" and "on." Its
action is simple: if it is off, then it will stay off forever. If it
is on, then it can detect a particle and go off. Later on we will
specialize to detection
of particle presence at a given location in space.
For a while let us be
general and assume that we have two Hilbert spaces ${\cal H}_{off},
{\cal H}_{on}$
and two Hamiltonians $H_{off}, H_{on}$. We also have a time dependent family of
operators $g_t :{\cal H}_{on}\rightarrow {\cal H_{off}}$ and let us denote
$\Lambda_t=g_t g_t^\star :{\cal H}_{on}\rightarrow{\cal H}_{on}$. 
According to
the theory presented in the previous section, with $g_{off,on}=
g_t$, $g_{on,off}=0$,
the master equation for the total system, i.e. for particle and detector,
reads:
\begin{eqnarray}
{\dot \rho}_{off}(t)&=&-i[H_{off},\rho_{off}(t)]+g_t \rho_{on}(t) g_t^\star 
\nonumber\\
{\dot \rho}_{on}(t)&=&-i[H_{on},\rho_{on}(t)]-{1\over2} 
\{\Lambda_t,\rho_{on}(t)\}.
\end{eqnarray}
Suppose at $t=0$ the detector is "on" and the particle state is 
$\psi (0)\in {\cal H}_{on}$, with $\Vert \psi (0)\Vert=1.$  
Then, according to the event generating algorithm
described in the previous section, probability of detection during
time interval $(0,t)$ is equal to $1-\Vert \exp (-iH_{on}t-{t\over2}
\Lambda_t )\; \psi (0)\Vert^2 .$

Let us now specialize and consider a detector of particle present at
a location $a$ in space (of $n$ dimensions). 
Our detector has a certain range of detection
and certain efficiency. We encode these detector characteristics 
in a gaussian function:
\be
g(x)=\kappa^{n/2}({\alpha\over\pi})^{3/2} \exp (-\alpha x^2).
\ee
If the detector is moving in space along some trajectory $a(t)$, and if
the detector characteristics are constant in time, then we put:
$g_t (x)=g(x-a(t))$. 
Let us suppose that the detector is off at $t=t_0$ and that the
particle wave function is $\psi_0(x)$. Then, according to the
algorithm described in the previous section, probability of detection
in the infinitesimal time interval $(t_0,t_0+\Delta t)$ equals 
$\int g_{t_0}^2 (x) | \psi_0 (x) |^2 dx \cdot \Delta t$. In the limit 
$\alpha\rightarrow\infty$, when $g^2_t (x)\rightarrow \kappa \delta (x-a(t))$
we get $\kappa |\psi_0 (a(t_0)) |^2 \cdot \Delta t$. Thus we recover the
usual Born interpretation, with the evident and necessary correction that 
the probability of detection is proportional to the length of exposure
time of the detector. \\
That simple formula holds only for short exposure times. For a prolonged 
detection, the formula becomes more involved, primarily
because of non-unitary evolution due to the presence of the detector. In that
case, numerical simulation is necessary. To
get an idea of what happens, let us consider a simplified case which
can be solved exactly. We will consider the ultra--relativistic Hamiltonian
$H=-i d/dx $ in space of one dimension. 
In that case the non-unitary evolution equation is easily solved:
\be
\psi (x,t) = e^{-{1\over2}\int_0^t \Lambda_s(x+s-t)} \psi(x-t,0).
\ee
In the limit $\alpha\rightarrow\infty$ when the detector shrinks to a point,
and assuming that this point is fixed in space $a(t)=a$, 
we obtain for the probability $p(t)$ of detecting the particle in the time
interval $(0,t)$:
\be 
p(t)=(1-e^{-\kappa}) \int_{a-t}^a |\psi(x,0)|^2 dx.
\ee
Intuitively this result is very clear. Our Hamiltonian describes
a particle moving to the right with velocity $c=1$, the shape of the
wave packet is preserved. Then $p(t)$ is equal to the probability that
the particle at $t=0$ was in a region of space that guaranteed passing
the detector, multiplied by the detector efficiency factor - in our case 
this factor is $1-e^{-\kappa}.$
\subsection{}
Let us consider now a relativistic Dirac particle.\cite{blaja96b} 
The main point in our approach
is to treat the relativistic case as a non-relativistic one, but replacing
time $t$ with "proper time" parameter $\tau$, and replacing Hamiltonian $H$
with Fock-Schwinger super-Hamiltonian. Explicitly:

We will take the standard representation of gamma matrices:
\be
\gamma^0=\pmatrix{I&0\cr0&-I\cr},\; \gamma^i=\pmatrix{0&\sigma^i\cr -\sigma^i&0\cr}
\ee
and define indefinite metric space by
\be
<\Psi , \Phi >=\int {\bar Psi}(x,t)\Phi(x,t) dx dt ,
\ee
where ${\bar\Psi}=\Psi^\dagger\gamma^0$ .
The Dirac matrices are Hermitian with respect to this scalar product, and so is the
Dirac operator:
\be
{\cal D}= i\gamma^\mu (\partial_\mu+ieA_\mu)-m .
\ee
Let  us consider now a particle position detector which, for simplicity, is at rest with respect to the
coordinate system. We associate with it the operator $G$ defined by
\be
(G\Psi)(x,t)={{I+\gamma_0}\over 2}g(x)\Psi(x,t),
\ee
where $g(x)$ is a positive, bell-like function centered over the detector
position. \footnote{Note that here,  as in the nonrelativistic case,
we assume that $g$ depends only on $x$ and not on $t$ - in the coordinate
system with respect the detector is at rest.}
It follows now that $G$ is positive, Hermitian with respect to
the indefinite metric scalar product, and the same holds for 
$\Lambda=G^2$. We postulate the following relativistic version of the PDP
algorithm:  

\begin{ralgorithm}
Suppose that at proper time  $\tau=0$ the 
system is described by a quantum state vector $\Psi_0$ and the counter 
is off: $\alpha=0$. 
Then choose a uniform random number $p\in [0,1]$, and proceed with
the continuous time evolution by solving the modified evolution
equation

\be
{\dot \Psi}_\tau =(-i{{{\cal D}^2}\over{2M}}-{1\over2}\Lambda )\psi_t
\label{eq:taupsi}
\ee
with the initial wave function $\Psi_0$ until $\tau=\tau_1$, where $\tau_1$
is determined by
$$\int_{0}^{\tau_1} (\Psi_\tau ,\Lambda \Psi_\tau ) d\tau = p.$$ At $\tau=\tau_1$
the counter clicks, that is its state changes from $\alpha=0$ to 
$\alpha=1$ and, at the same time, the state vector jumps:  
$$\Psi_{\tau_1}\rightarrow\Psi_1=G\Psi_{\tau_1}/
<\Psi_{\tau_1},G\Psi_{\tau_1}>.$$ 
The evolution starts now again and it obeys the standard unitary
Schr\"odinger equation with the Hamiltonian $H$.
\end{ralgorithm}

The above prescription is not the only one possible. But it has one
very important property: the algorithm is {\sl independent} of any local
observer. It is, in fact, a somewhat strange algorithm - it
works {\sl as if it was quite natural for Nature to be working in more than
four space-time dimensions.} 
\subsection{}
In EEQT it is possible to model a simulataneous measurement of
several non-commuting observables. And example would be a simultaneous
measurement of the same component of position an momentum. This case,
however, has not yet been studied - because of its computational
difficulties. A simpler problem, namely that of a simultaneous
measurement of several spin projections leads to chaotic behavior
and fractal structure on the space of pure states. Following
the discussion given in \cite{jadifs} let us couple a spin $1/2$
quantum system to four yes-no polarizers corresponding to spin
directions ${\bf n_i}$, $i=0,1,2,3$, arranged at the vertices of
a regular tetrahedron. Choosing the same coupling structure
$\kappa$ for all four polarizers the model leads to a homogeneous
(in time) Poisson process on the sphere $S^2$ of norm $1$ quantum spin states.
The process is a non-linear version of Barnsley's iterated function
system \cite{barn} and can be described as follows:\\
for $i=0,1,2,3$ let $a_i$ be the $2$ by $2$ matrices
 $a_i = \frac{I+\alpha {\bf n}_i\cdot{\bf s}}{2}$, where ${\bf s}$ are the Pauli
matrices, and let $A_i$ be
the four operators acting on $S^2$ by $\phi\mapsto a_i\phi/\Vert a_i\phi\Vert .$
These operators play the role of Barnsley's affine transformations. To each
transformation there is associated probability 
$p_i=\frac
{1+\alpha^2+2\alpha {\bf n}_i\cdot{\bf r}}
{4(1+alpha^2)}$, 
where ${\bf r}$ is the radius-vector of the
actual point on the spehere, that is to be transformed. Iteration leads to
a self--similar structure, with sensitive dependence on the initial state
and on the value of the coupling constant. Numerical simulation shows that
when $\alpha$ decreases from $0.95$ to $0.75$, Hausdorff dimension of the 
limit set increases from $0.5$ to $1.3$. Fig. 1 shows a typical picture - here for $\alpha=0.73 .$ For details see \cite{jadjas}.
\section{EEQT: FAQ}
In this section we answer a series of questions and objections that
are being raised concerning the formalism and implications of EEQT.
 
\begin{enumerate}
\item {\sl Isn't it so that EEQT is a step backward toward classical mechanics
that we all know are inadequate?}\\
EEQT is based on a simple thesis: not all is "quantum" and there are things in this universe
that are NOT described by a quantum wave function. Example, going to an extreme: one such case is the wave function itself. Physicists talk about first and second quantization.  Sometimes, with considerable embarassment, a third quantization is considered.  But that is usually the end of that. Even the most orthodox quantum physicist controls at some point his "quantize
everything" urge - otherwise he would have to "quantize his quantizations" ad infinitum, never beeing able to communicate his results to his colleagues. The part of our reality
that is not and must not be "quantized" deserves a separate name. In EEQT we are using the term "classical." This term, as we use it, must be understood in a special, more-general-than-usually-assumed way. "Classical" is not the same as "mechanical."  Neither is it the same as "mechanically deterministic."  When we say "classical" - it means "outside of the restricted mathematical formalism of Hilbert spaces, linear operators and linear evolutions." It also means: the value of the "Planck constant" does not govern classical parameters. Instead, in a future theory, the value of the Planck constant will be explained in terms of a "non-quantum" paradigm. More on this subject in Section 6.\\
%%%%%%%%%%
\item {\sl The mathematical framework of EEQT seems to be routine.}\\
EEQT is based on a little known mathematical theory of piecewise 
deterministic processes. It is impossible to discuss it rigorously 
without applying this theory. In fact, it is impossible to discuss
any variation of a quantum theory that incorporates "events"
as an inhomogeneous Poisson process without using PDP's.
GRW avoids this requirement only because it assumes a 
homogeneous Poisson process. And, it is clear that any attempt to 
incorporate Einstein's relativity or non-uniformly accelerated 
observers would lead to inhomogeneous processes. In fact, as shown
in \cite{ja94b,ja94c} GRW can be considered as a particular, degenerate, case
of EEQT.\\
There can be no understanding of what EEQT is about without
understanding the rudiments of PDP theory; and there are
only two or three books dealing with this theory. The mathematics
involved are NOT routine. In fact,it requires a very clever
application of PDP. This is due to the fact that, in quantum theory, 
we have at our disposal not the full algebra of functions on pure 
states but only small subsets of bilinear functions - given by 
expectation values of linear operators.
\item {\sl EEQT is too abstract for immediate applications to any
concrete problems.}\\
%%%%%%%%%%%%%
EEQT has been applied to several problems, the most developed 
being  its application to tunneling time.\cite{muga97,rusch} There it gives predictions 
that can be tested experimentally and compared with those 
stemming from other approaches. In this respect, orthodox 
quantum theory gives no predictions at all. Orthodox quantum 
theory is helpless when it comes to predicting timing of events. The 
classic paper by Wigner that tried to deal with the subject is 
inconclusive and has errors. The most evident future application
of EEQT that we envisage relate to quantum computations, where
EEQT formalism will provide interface between quantum and classical
computing units.\\
%%%%%%%%%%%%%%%%
\item {\sl I doubt if EEQT is sufficiently important that its properties should be of 
wide interest.}\\
As noted above, EEQT gave predictions concerning tunnling times. 
These predictions may prove to be right or wrong. If they
prove to be wrong - then it will mean that EEQT is wrong. If they
prove to be right - then it will mean that EEQT is better than any
other competing approach. WE believe that any theory that
is based on a healthy and rigorous math, reduces to known
theories in a certain domain,AND predicts more in other
domains SHOULD be of wide interest. In EEQT, contrary to the 
standard quantum theory, there is no need to invoke "external
observers." ALL is in the equations. EEQT, in contrast to
the orthodox QT,  provides its own interpretation. We believe 
that these factors make it of interest to a wide audience.\\     
%%%%%%%%%%%%%%%%
\item {\sl EEQT is presented as a solution to the quantum measurement problem. As such
it must be compared with the other proposed solutions. The authors mention
two alternatives: the spontaneous localization idea of Ghirardi, Rimini,
and Weber (GRW) -cf \cite{grw} - and hidden variables, e.g. Bohm's theory - cf 
\cite{bohm}. 
Now on the surface
these two alternatives seem so vastly superior and so much better developed
than EEQT that it is hard to understand why anyone should pay much attention to EEQT.}\\
GRW has nothing to say about tunnelling time problems. Bohm's 
theory predictions are different from those of EEQT. Bohm's theory
is also more than 40 years old. It takes time to develop a theory.
We are not presenting a fully developed theory. We are presenting
a theory that is BEING developed, but even at an incomplete stage,   EEQT gives new predictions that can be tested experimentally. That is why we believe EEQT requires attention - even if only to disprove it - if possible.\\
%%%%%%%%%%%%
\item {\sl While the authors give no indication of why EEQT should be regarded 
 as improving in any way on GRW, they do say that their formalism "avoids 
 introducing other hidden variables beyond the wave function 
itself." But this is not true, except in a sense for which the same thing
 could be said for any HV theory.}\\
Hidden variable theories use microscopic hidden variables that are 
"hidden" indeed from our observations! EEQT deals with
classical variables that can be observed. In fact, it states that
these are the ONLY variables that can be observed. 
Classical variables of EEQT are a direct counterpart of physics
on the other side of the Heisenberg-von Neumann cut.\\
\item {\sl In EEQT, quantum mechanics is supplemented
by a "classical system" (an apparatus?) given by an Abelian algebra of
observables that also commute with all quantum observables. The spectrum of
this algebra corresponds precisely to the possible values of the classical
variables.  Now in fact, any (hidden) variables in addition to the wave
function could also be similarly regarded as corresponding to the spectrum of
the center of an algebra of observables containing the quantum algebra.}\\
This is not true. Hidden variable theories of Bohm and of Bell are 
incompatible with linearity. They can not be formulated in algebraic 
terms at all. The statement that their hidden variables could be 
considered as corresponding to the spectrum of the center of an algebra of observables containing the quantum algebra is incorrect. It is based on misconception. EEQT is compatible with linearity. There is a reflection of
this fact in the following: In hidden variable
theories there is NO back action of classical variables on the wave
function. In EEQT there is such an action. Linearity imposes
the need for such a reciprocical action.\\
%%%%%%%%%%%%%%%%%%
\item {\sl In his celebrated analysis of the quantum measurement problem,
"Against Measurement," John Bell indicates that to make sense of the usual
mumbo jumbo one must assume either that (i) in addition to the wave
function psi of a system one must also have variables X describing the
classical configuration of the apparatus or (ii) one must abrogate the
Schr\"odinger evolution during measurement, replacing it by some sort of
collapse dynamics. EEQT is a theory combining (i) and (ii): there are
additional classical variables and because of the interaction between these
variables and the quantum degrees of freedom, the evolution is not exactly
the Schr\"odinger evolution and leads to collapses in measurement
situations.}\\
This is true.
\item {\sl Now Bell criticizes (i) and (ii) because they ascribe a special
fundamental role to measurement, which seems implausible and makes
vagueness unavoidable.}\\
In EEQT we distinguish between a measurement and an 
experiment. Our universe can be considered as being "an
experiment." This is in total agreement with Bell.\\ 
\item {\sl He then goes on to suggest two ways to overcome this
difficulty: by not limiting X to macroscopic variables one arrives at
Bohm's theory and by introducing a suitable microscopic collapse mechanism
at GRW (as the simplest possibility).}\\
This is what Bell knew at the time of writing his papers. EEQT did
not exist at this time. There are certainly more options
available. EEQT shows that there are such options. But, as stated, 
EEQT is not yet a complete theory. It is semi-phenomenological. 
Its aim is to find the ultimate classical parameters without stating
a-priori restrictions on their nature. They may prove to be
related to gravity a'la GRW and Penrose; they can be related
to consciousness a'la Stapp and Penrose-Hameroff; they can
be related to new kind of fields that are yet to be discovered.
John Bell was open minded. EEQT is open-minded as well.\\
%%%%%%%%%%%
\item {\sl We've made a great deal
of progress in the past few decades, progress that is not reflected in EEQT.} \\
None of this progress helps us to better understand such a 
simple phenomenon as predicting tunelling time for an individual particle. Much of the so called
"progress" leads to no new predictions. EEQT does.\\
\item
{\sl The name "Event Enhanced Quantum Theory" is misleading.}\\
As we have stated: "EEQT is the minimal extension of orthodox 
quantum theory that allows for events." It DOES enhance 
quantum theory by adding the new terms to the Liouville equation. 
When the coupling constant is small, events are rare and EEQT 
reduces to orthodox quantum theory. Thus it IS an enhancement.\\
%%%%%%%%%%%%%%%%%
\item {\sl The possibility is thus opened for experimental discrimination
between the two theories. Unfortunately, EEQT is formulated in too abstract and 
schematic a manner to permit any such discrimination. }\\
We agree that what is lacking is a textbook presentation of EEQT,
with a thorough presentation of its experimental consequences
and its relation to the orthodox QT. Writing such a textbook
is presently being considered.\\
\item {\sl It seems almost as if the
coupling of classical to quantum degrees of freedom, given by the matrix
$g_{\alpha,\beta}$ of linear operators and defining EEQT for the case at hand,
is to be just so chosen as to reproduce the quantum predictions for the
measurements under discussion. }\\
It is such that it reproduces those quantum predictions that 
have already been tested, but it also gives new predictions,
about which quantum theory is silent, concerned with timing of the events and with the back action of the classical variables on the wave function. Any new, useful theory must be built in such a way that it is in agreement with
the succesful aspects of the old one.  EEQT is no exception in this 
respect.  The point is that it differs from OQT in predicting more, and in
predicting corrections to OQT predictions. \\
%%%%%%%%%%%%%%%%
\item {\sl If the authors could provide a more general formulation of their theory, first by being clear about how the line is to be drawn between classical degrees of freedom and quantum ones; how the autonomy of the classical degrees of freedom fits with the fact, presumably accepted by the authors, that classical degrees of freedom are built out of quantum degrees of freedom...}\\
No, the authors do NOT presume this! Such a presumption is not
justified by experiments. Experiments show that we are living
in the world of FACTS, not the world of POSSIBILITIES. The authors
do presume, that THERE IS a classical part of the universe that is
not reducible to quantum degrees of freedom. Assuming that all
must be quantum is similar to believing that the Sun revolves 
around the Earth. Without adequate knowledge, this seems to be 
observably so.  EEQT is in agreement with all observable facts in 
at least the same degree as pure quantum theory is. But EEQT 
accomodates a knowledge base which accounts for events while 
quantum theory can't.\\
%%%%%%%%%%%%
\item  {\sl ... and then by providing some
general specification of the interaction between classical and quantum
degrees of freedom, analogous to specifying that electrons are governed by the
Coulomb interaction or by QED, we would thereby have an alternative to
quantum theory making perhaps dramatically different predictions from that
theory. This might well be worth our consideration. }\\
When such a theory is finished and ready - it will certainly deserve 
a Nobel Prize!  EEQT is not yet at this stage. Nor are any
of the competing theories. However, we are working toward this
end.
\end{enumerate}
%%%%%%%%%%%%%%%

\section{Quantum Future} 
First of all EEQT itself needs to be further developed in order to provide a theoretical and computational answer to the needs of modern quantum engineering
and technology. We need to include classical systems with infinite
numbers of degrees of freedom - like electromagnetic and/or gravitational
field.  We also need to include infinite quantum systems so as
to understand and simulate Bose gas and its phase transitions. But, there
are also steps that must be taken far beyond the paradigm of EEQT.\\    
As we have emphasized so many times: quantum theory has yet
to be understood in terms of a "non-linear classical theory." But what we mean by a "classical theory" is something much more advanced and more
general than clasical mechanics or classical field theory. By a "classical theory" we mean first of all, a theory that is not based
on probabilities from the start; a theory in which probabilities appear  at a
later stage, derived from the theory; derived perhaps, in a necessary way. Such a theory must not only  specify the mathematical objects and their relations but also make "predictions".\\
A classical theory is, in particular,  a theory in which discrete events can happen, events that are "objective."  They may be events that affect mainly the physical stratum, but they may be also "mental 
events," changes of states of "consciousness;" whatever they are, even if they
are concerned with branchings of universes, - they DO HAPPEN.\\
Such a theory, encompassing the quantum theory as we know it today,
does not yet exist, but we can envisage its possible shape. Nothing serves better than an example, so let us give here an example of how
such a theory can be construed.\\
Imagine a theory developed a'la Einstein's unified field theory, but
with variable metric signature, possibly with a complex causal web, multiple
Einstein--Rosen \cite{e-r} (see also \cite{ajad82}) bridges,
time loops, nondifferrentiability, fractal structure, all of that additionally complicated by variable dimensionality of space and of "time" (cf \cite{coja88}. Imagine such a theory to be able to accomodate all the four fundamental
forces known to us, but it also involves an extra field, which is different
from physical fields, and which is non-local in the sense that it does not
survive taking a macro-average-limit when causal  space-time 
structure of Einstein's general relativity is recovered. This extra field
would be a place for "thought forms" and our theory would couple these
thought forms to more "physical" levels of reality. The very concept of
"time" would arise only in one particular limiting structure. Nothing 
prevents a theory of such a type to have mathematical structures rich enough to accomodate consciousness and mind. Due to its complexity,
density of time-loops, bubble-like causal structure, making predictions
in such theory is possible only by applying probability -
as it is the case with systems evolving according to deterministic
but chaotic dynamics. Quantum wave function would emerge in such a
theory as an effective way of  predicting. That type of explanation
of quantum indeterminism was postulated long ago by I. J. Good. In 
\cite{good} he speculated that 
quantum indeterminism can be understood if we admit that individual quantum events that actually happen here and now have future
advanced causes as well has past retarded causes.\footnote{More than thirty
years later time travel 
is in the field of active research of NASA - cf. http://www.lerc.nasa.gov/WWW/bpp/}\\
A glimpse of such a thought
may have occured to Alan Turing when he wrote his famous, already quoted sentence: "prediction must be linear, description must be nonlinear."\\   
A theory going partially in this direction is being developed by L. Nottale
\cite{nottale} who states the observation, attributed to Feynman, that "the typical paths of quantum mechanical particles are continuous non-differentiable." Then he continues his development of fractal
space-time ideas to conclude with: "The quantum behavior becomes, in this theory, 
a manifestation of the fractal geometry of space-time, in the same way gravitation is, in Einstein's theory of general (motion-)relativity, a manifestation of the curvature of space--time.\\
Another step in this direction comes from the work of Russian scientists - which started with the most original ideas of N.A. Kozyrev \cite{kozyrev}
and A. Sakharov \cite{sakharov}, and is being developed in Moscow, the Urals and Siberia, see e.g.  \cite{frolov, guts, eganova}.
\section{Summary and conclusions}
In Section 3 we have given just a few examples of simple models based on EEQT; more can
be found elsewhere. In \cite{blaja93b} quantum Zeno effect is discussed within the framework
of EEQT. A quantum system is "observed" - that is coupled in an appropriate way - to a classical system. The intensity of observation is mathematically modelled by the
value of the coupling constant. We find that indeed, with the increase of the
coupling constant, the Hamiltonian part of the evolution effectively stops.\\
In \cite{blaja94c} we examined the EPR paradox within EEQT, with the result that EPR 
phenomenon alone can not be used for a superluminal signalling.\\ 
In \cite{ja94a,blaja93d} we discussed the problem of whether the quantum state
itself can be determined by a measurement (as defined within EEQT)\\
In \cite{blaja93c} we
applied EEQT to a SQUID--tank model, where a classical system has as its 
manifold of pure states, the phase space of a radio-frquency oscillator. It is
interesting that in this case classical events are characterized by discontinuous
changes of velocity, while the position is changing in a continuous way. The back
action of the  quantum circuit on the classical one leads to new terms in
semi-phenomenological evolution equation that, in principle, can be tested
experimentally.\\ Some mathematical problems arising in our models have been discussed
by Olkiewicz in \cite{rolek}, while in \cite{blajaol} we have examined in detail
the relevant probabilistic aspects of the piecewise deterministic Markov process
governing the behaviour of individual systems.\\ Quantum time--of--arival observables, 
its non-linearity and dependence on the effectiveness of the detectors have been
discussed in \cite{blaja95f}, while in \cite{blaja95a} we have shown that Born's
probabilistic interpretation of quantum wave function follows, in a special limit,
from our detector model. An entropy generating  fuzzy clock is discussed in
\cite{blaja95c}\\ Algorithm for cloud-chamber particle tracks formation, 
resulting from EEQT have been developed in \cite{ja94b,ja94c}.\\ 
Some projects we have started
are still in a state of incompletion for lack of time. One such project is deriving EEQT from quantum electrodynamics, where the classical parameter enters naturally as the index
of inequivalent non-Fock infrared representations. We believe that using
infinite tensor product representations of quantum systems with an infinite number
of degrees of freedom, we will arrive naturally at our $g_\alpha\beta$ operators
relating to Hilbert spaces of inequivalent representations of CCR\/CAR. \\We also
started, but did not finish, modelling of a coupling of a quantum particle to 
a classical (Newton) gravitational potential. The general idea is simple: a quantum particle,
having a mass, must back-react on the gravitational potential. The point is, however,
to model it, in a natural and possibly unique way, via Louville equation of the type demanded by EEQT. This would be only a first step towards a more ambitious project: coupling
of Quantum Electrodynamics to classical (relativistic) gravity, with -- how else - a hope that
the back--action will smooth out divergencies of QED.\\ Some of the future project are
rather straightforward - here belongs the further study of chaos induced by quantum
measurement. We encourage all interested readers to contact us - we will try to help
as much as we can.
\vskip10pt
\noindent
{\bf Acknowledgements}\\
One of us (A.J) would like to thank L.K-J for invaluable help.

\newpage
\begin{figure}[t]
\caption{Barnsley iterated function system on the unit sphere of a two-dimensional Hilbert space - resulting from simultaneous monitoring of four non-commuting components of a spin.
The four spin components are situated at the corners of a regular tetrahedron. View from
the North Pole - which is one of the tetrahedron's corners. }
\label{fig:fig1}
\end{figure}

\vskip10pt
\noindent

\begin{thebibliography}{99}
\bibitem{born} Blanchard, Ph., and Jadczyk, A. Ed. {\sl Quantum Future}, Proceedings of the X-th Max Born Symposium, to appear, Springer 1999
\bibitem{stapp} Stapp, H.P.: \lq Quantum Ontology and Mind-Matter Synthesis.\rq , also
available from http://www-physics.lbl.gov/\~stapp/stappfiles.html 
\bibitem{good} Good, I.J.: {\sl The Scientist Speculates}, Heineman, 1962
\bibitem{schrod55} Schr\"odinger, E.: \lqq The Philosophy of Experiment\rqq,
{\sl Il Nuovo Cimento, }{\bf 1} (1955), 5-15
\bibitem{ja94b}  Jadczyk, A.: \lq Particle Tracks, Events and Quantum
Theory\rq,  Progr. Theor. Phys. {\bf 93} (1995) 631--646, available from
hep-th/9407157
\bibitem{ja94c} Jadczyk, A.: \lq On Quantum Jumps, Events and Spontaneous
Localization Models\rq, {\em Found. Phys.} {\bf 25} 1995)743--762,
available from hep-th/9408020
\bibitem{bell90}  Bell,  J. : \lqq Against measurement\rqq, in
{\sl Sixty-Two Years of Uncertainty. Historical, Philosophical and
Physical Inquiries into the Foundations of Quantum Mechanics}, Proceedings
of a NATO Advanced Study Institute, August 5-15, Erice, Ed. Arthur I. Miller,
NATO ASI Series B vol. 226 , Plenum Press, New York 1990
\bibitem{bell89} Bell,  J. : \lqq Towards an exact quantum mechanics\rqq,  in
{\sl Themes in Contemporary Physics II.  Essays in honor of Julian
Schwinger's 70th birthday},  Deser,  S. ,  and Finkelstein,  R. J.  Ed. ,
World Scientific,  Singapore 1989
\bibitem{haag90a}Haag, R.: \lq Irreversibility introduced on a fundamental level\rq ,
{\sl Commun. Math. Phys.} {\bf 123} (1990) 245-251
\bibitem{haag90b}Haag, R.: \lq Thought of the Synthesis of Quantum Physics and
General Relativity and the Role of Space--time\rq , {\sl Nucl. Phys.} {\bf B18}
(1990) 135--140
\bibitem{haag96a} Haag, R.: \lq An Evolutionary Picture for Quantum Physics\rq ,
{\sl Commun. Math. Phys.} {\bf 180} (1996) 733-743
\bibitem{haag96b} Haag, R.: {\sl Local Quantum Physics}\ 2nd rev. and enlarged
ed. (1996), Ch. VII: {\em Principles and lessons of Quantum Physics. A Review of 
Interpretations, Mathematical Formalism and Perspective.}
\bibitem{haag98} Haag, R.: \lq Objects, Events and Localization \rq ,
in \cite{born}
\bibitem{deutsch} Deutsch, D.: {\sl The Fabric of Reality}, Allen Lane The Penguin
Press, 1997
\bibitem{wolf} Wolf, F.A.: {\sl Parallel Universes}, Touchstone, New York 1990
\bibitem{blaja93a} Blanchard,  Ph.  and Jadczyk,  A.: \lqq On the interaction
between classical and quantum systems\rqq, {\sl Phys. Lett. }{\bf A
175} (1993), 157--164
\bibitem{blaja95a} Blanchard, Ph., and A. Jadczyk.: \lq
Event--Enhanced--Quantum Theory and Piecewise Deterministic Dynamics\rq,
{\em Ann. der Physik} {\bf 4} (1995) 583--599,
\bibitem{blaja95e} Blanchard, Ph., and Jadczyk, A.: \lqq
Events and Piecewise Deterministic Dynamics in
  Event-Enhanced Quantum Theory\rqq , {\em Phys. Lett.} {\bf A203} (1995),
260--266
\bibitem{blaja95f} Blanchard, Ph., Jadczyk, A.: \lq Time of Events in Quantum
Theory\rq ,  {\sl Helv. Phys. Acta} {\bf 69} (1996) 613--635, also
available as quant-ph/9602010,
\bibitem{haroche} Haroche, S.: \lq Observing the decoherence of the meter in a 
measurement: a variation on Schrodinger's cat experiment\rq , in \cite{born}
\bibitem{walther} Walther. H.: \lq Quantum Optics of a Single Atom\rq ,
in \cite{born}
\bibitem{qubit0} Ekert, A.: Web page of Centre of Quantum Computation of the University of Oxford, URL address http://www.qubit.org/intros/crypt.html
\bibitem{qubit} Benjamin, S., and Ekert, A.: as above, URL address http://www.qubit.org/Intros\_Tuts.html
\bibitem{muga97} Palao, J. P.,  Muga, J.G., Brouard, S.  and Jadczyk, A.:
\lq Barrier traversal times using a phenomenological track formation
model\rq , {\sl Phys. Lett.} {\bf A 233} (1997) 227-232
\bibitem{rusch} Rushhaupt, A.: \lq Simulations of Barrier Traversal and Reflection
Times based on EEQT\rq, {\sl Phys. Lett} {\bf A 250}, (Dec. 1998?), 249--256 (Dec. 1998?)
\bibitem{jakol95} Jadczyk, A., Kondrat, G. and Olkiewicz, R. : \lq On
uniqueness of the jump process in quantum measurement theory\rq ,
{\sl J. Phys.} {\bf  A 30} (1996) 1-18, available from quant-ph/9512002 
\bibitem{koss} Gorini,  V. ,  Kossakowski,  A.  and Sudarshan, 
E. C. G. : {\em Completely positive dynamical semigroups of N-level
systems},  {\sl J.  Math.  Phys.} {\bf 17} (1976), 821-825
\bibitem{lin} Lindblad,  G. : {\em On the Generators of Quantum Mechanical
Semigroups}, {\sl Comm. Math. Phys. } {\bf 48} (1976), 119-130
\bibitem{arv} Arveson,  W.  B.  : "Subalgebras of $C^{\star}$--algebras", 
{\sl Acta Math. }{\bf 123} (1969), 141--224
\bibitem{chr} Christensen,  E.  and Evans,  D. : {\em Cohomology of
operator algebras and quantum dynamical semigroups},  {\sl J.  London. 
Math.  Soc. }  {\bf 20} (1978), 358-368
\bibitem{blaja96b} \lq Relativistic Quantum Events\rq , {\sl Found. Phys.}
{\bf 26} (1996) 1669-1681
\bibitem{barn} Barnsley, M.F.: {\sl Fractals Everywhere}, Academic Press, 
Boston 1988
\bibitem{jadifs} Jadczyk, A.: \lq IFS Signatures of Quantum States\rq, IFT Uni Wroclaw,
internal report, September 1993.
\bibitem{jadjas} Jastrzebski, G.: Ph.D. thesis (in Polish); Jadczyk, A., and Jastrzebski, 
G.: \lq Chaos from Quantum Mesurements\rq , to be published.
\bibitem{grw} Ghirardi, Gian-Carlo - in this volume, for a recent popular account see 
also internet URL address http://www.newscientist.co.uk/ns/970426/features.html
\bibitem{bohm} Bohm, D., and Hiley, B.:{\sl The Undivided Universe},
Routledge, London 1994
\bibitem{davmha1} Davis, M. H. A. : {\em Lectures on Stochastic
Control and
Nonlinear Filtering}, Tata Institute of Fundamental Research, Springer
Verlag, Berlin 1984
\bibitem{davmha2} Davis, M. H. A. : {\em Markov models and optimization},
Monographs on Statistics and Applied Probability,  Chapman and Hall,  
London 1993 
\bibitem{rolek}
Olkiewicz, R. : \lq Some mathematical problems related to classical-quantum
interactions\rq ,\\ {\sl Rev. Math. Phys.} {\bf 9} (1997), 719-747 
\bibitem{e-r} Einstein, A., and Rosen, N.: {\sl Phys. Rev.} {\bf 48}, (1935),
73--77 
\bibitem{ajad82} Jadczyk, A.: {\sl Vanishing Vierbein in Gauge Theories of
Gravitation}, preprint ITF Uni. Goettingen, August 1982, available from
http://www.geocities.com/~lark22/jadpub.html
\bibitem{coja88} Coquearaux, R., and Jadczyk, A.: {\sl Riemannian Geometry Fiber Bundles Kaluza-Klein Theories and All that ... ., }, World Scientific, Lecture Notes in Phys. {\bf 16},  (1988)
\bibitem{nottale} Nottale, P.:\lq Scale Relativity and Fractal Space--Time: Applications to
Quantum Physics, Cosmology and Chaotic Systems\rq , {\sl Chaos, Solitons and Fractals}
{\bf 7} (1996) 877--938 
\bibitem{kozyrev} {\sl On the Way to Understanding the Time Phenomenon. The
Construction of Time in Natural Science. Part 2: The "Active" Properties
of Time According to N.A. Kozyrev.}, World Scientific, Series on Advances
in Mathematics for Applied Sciences - {\bf Vol. 3}, (1996)
\bibitem{sakharov} Sakharov, A. D. \lq Vacuum Quantum  Fluctuations in Curved Space
                      and the Theory of Gravitation\rq , {\sl Dokl. Akad. Nauk.
                      SSSR} (Sov.  Phys.  - Dokl.) {\bf 12}  (1968) 1040
\bibitem{frolov} Frolov, V.P., and Novikov, I.D.: \lq Physical
effects in wormholes and time machine\rq , {\sl Phys. Rev.} {\bf D42} (1990), 
1057--1065 
\bibitem{guts} Guts, A.K.: \lq Time machine as four--dimensional wormhole\rq , available from gr-qc/
9612064; cf. also references therein.
\bibitem{eganova} Eganova, I.A.:{\sl The World of Events Reality}, book, to appear 1999 
\bibitem{blaja93b} Blanchard, Ph. and Jadczyk, A.: \lq Strongly coupled
quantum and classical systems and Zeno's effect\rq, {\em
Phys. Lett. }{\bf A 183} (1993) 272--276

\bibitem{blaja94c} Blanchard, Ph. and Jadczyk, A.: 
\lq Event-Enhanced Formalism
 of Quantum Theory or Columbus Solution to the Quantum Measurement 
Problem\rq,
in {\sl Quantum Communications and Measurement},
Edited by V.P. Belavkin, Plenum Press, New York 1995, pp. 223--233 
\bibitem{ja94a}  Jadczyk, A.: \lq Topics in Quantum Dynamics\rq, 
in {\em Infinite Dimensional Geometry, Noncommutative Geometry,
Operator Algebras and Fundamental Interactions}, pp. 59--93, ed. R.Coquereaux et al.,
World Scientific, Singapore 1995, hep--th 9406204
\bibitem{blaja93d}  Blanchard,  Ph.  and Jadczyk,  A.: \lq Classical and
quantum
intertwine\rq,  in {\sl Proceedings of the Symposium on Foundations of
Modern Physics},  Cologne,  June 1993,  Ed.  P. Bush et al.,  World
Scientific  (1993), pp 65-76,  hep-th/9309112
\bibitem{blaja93c} Blanchard, Ph. and Jadczyk, A.:  \lqq How and
When Quantum Phenomena Become Real\rqq ,  in Proc.
Third Max Born Symp.  {\em Stochasticity and Quantum Chaos},
Sobotka 1993,  pp. 13--31, Eds.  Z.  Haba et all. ,  Kluwer Publ. 1994
\bibitem{blaja95c} Blanchard, Ph. and Jadczyk, A.:
\lq Quantum Mechanics with Event Dynamics\rq,
{\sl Rep.Math.Phys.} {\bf 36} (1995) 235--244
\bibitem{blajaol} Blanchard, Ph., Jadczyk, A., and Olkiewicz, R.:
\lq The Piecewise Deterministic Process Associated to EEQT\rq, Preprint
Uni. Bielefeld Forschungszentrum BiBoS, Nr. 802/4/98
\end{thebibliography}
\end{document}